\documentstyle[11pt,paspconf,psfig]{article}

\begin{document}

\title{
Frequency dependent radio structure of the gravitational lens system
B0218+357
}

\author{Alok R. Patnaik and Richard W. Porcas}
\affil{Max-Planck-Institut f\"ur Radioastronomie,
Auf dem H\"ugel 69, D 53121 Bonn, Germany.
}

\begin{abstract}
We present multi-frequency radio continuum VLBI observations of the
gravitational lens
system B0218+357 carried out using a global VLBI network and the VLBA.
The source has been observed with resolutions from 0.2~mas to 5~mas
and displays interesting structure. The spectral properties of various
components show that the lensed object is a standard flat spectrum radio
source which has many self-absorbed components. Based on the flux
ratio of the 
lensed images as a function of frequency we propose a simple model for
the background radio source.

\end{abstract}




\section{Introduction}

The radio source B0218+357 has been identified as a gravitationally
lensed system (Patnaik et al. 1993). The source, which has a
core-halo structure at low frequency (1.4~GHz) and low resolution
(5~arcsec), consists of two compact flat-spectrum components, A and B,
separated by 335 milliarcsec. The weaker of the two, B,
is surrounded by a faint Einstein ring of similar diameter. The
spectral and polarisation characteristics of the components show them to
be the lensed images of a single flat spectrum `core'. The lensing
galaxy, suggested to be a spiral galaxy, has been observed by the HST
(Jackson et al. 1997). It has a redshift of 0.6847
(Browne et al. 1993). Atomic hydrogen and many molecular species
have been detected in absorption against the background source (Carilli et
al. 1993, Wiklind \& Combes 1995, Menten \& Reid 1996). The redshift
of the background object is suggested to be 0.96 (Lawrence 1996). 

In this contribution we explore the properties of the radio source
from our multi-frequency VLBI observations. We give a summary of our
observations and discuss the results in the
context of a flat spectrum radio source. We note that the
overall radio spectrum of B0218+357 is flat between 365~MHz and 43~GHz
i.e. its flux density is constant within a
factor of two between these frequencies. 

\section{Observations}

B0218+357 has been observed at 1.7, 5, 15, 22 and 43~GHz over the last
several years in order to understand its small--scale structure and
thereby derive constraints for the lens models. The 1.7 and 5~GHz
observations were carried out in 1992 with a global array using the MK3 VLBI
system and the VLBA was used for the higher frequency observations.
About 10 telescopes were used in each of the observing sessions.  The
global array data were correlated at the MPIfR, Bonn.
The data at 15, 22 and 43~GHz were taken
using the VLBA with a bandwidth of 64~MHz.
The data were kept in 8~MHz channels to avoid bandwidth smearing.
The data were analysed using the NRAO {\sc AIPS} software package.

\section{Results}

The 1.7, 5 and 15~GHz maps are published by Porcas
\& Patnaik (1996a), another 15~GHz map and 22 and 43~GHz maps are published by
Porcas \& Patnaik (1996b).  In this contribution we show a
15~GHz map (Patnaik, Porcas \& Browne 1995) to identify the different
components (Fig. 1).

\begin{figure*}
\raisebox{-3.2cm}
{\begin{minipage}{0.3cm}
\mbox{}
\parbox{0.3cm}{}
\end{minipage}}
\begin{minipage}{5.0cm}
\mbox{}
{\psfig{figure=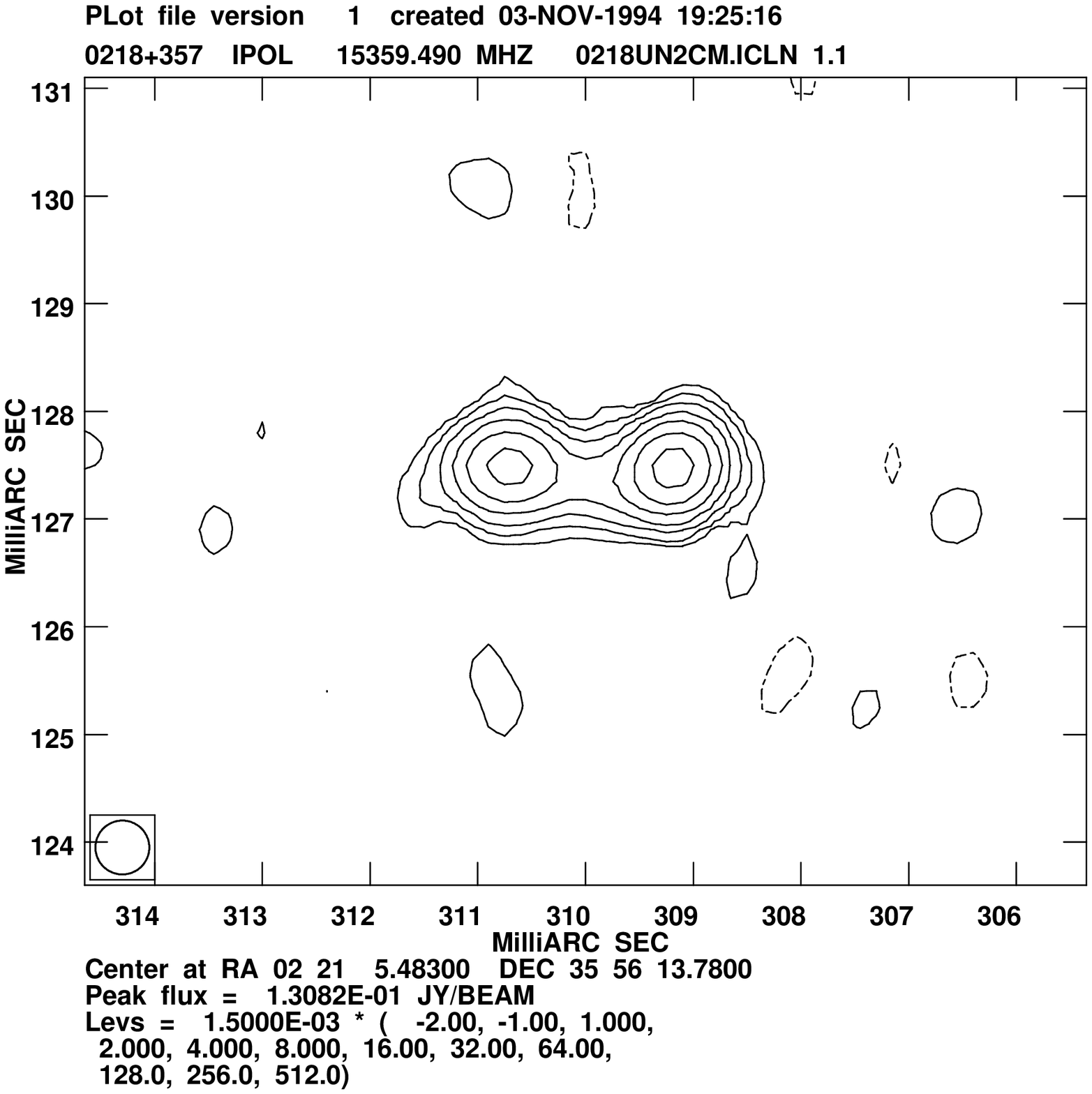,width=5.0cm,rheight=5.0cm,rwidth=5.0cm}}
\centering
\end{minipage}
\hspace{0.5cm}
\raisebox{-3.2cm}
{\begin{minipage}{0.3cm}
\mbox{}
\parbox{0.3cm}{}
\end{minipage}}
\begin{minipage}{5.0cm}
\mbox{}
{\psfig{figure=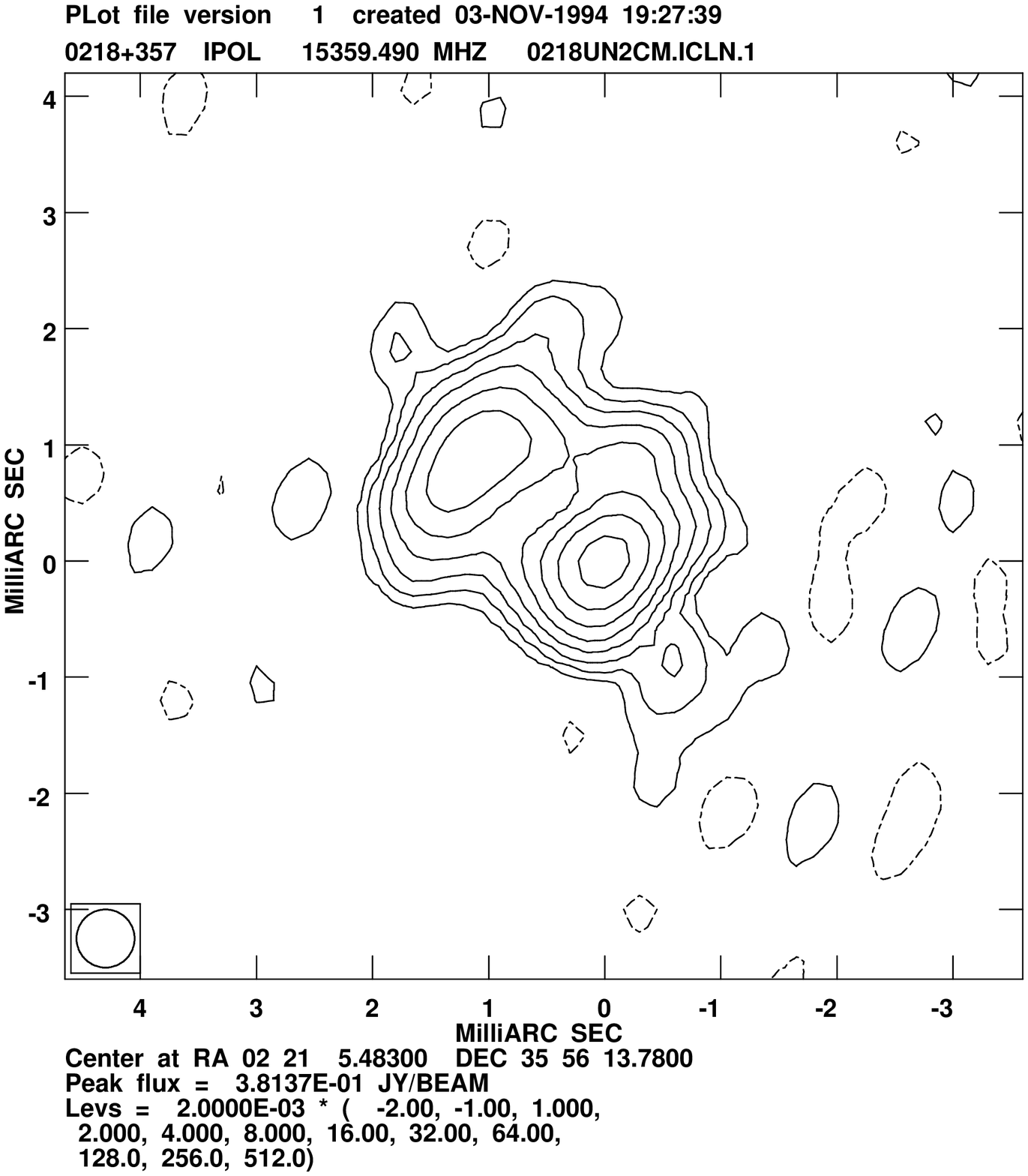,width=5cm,rwidth=5cm,rheight=5.0cm}}
\centering
\end{minipage}
\caption{
15~GHz VLBA maps of B0218+357 
images A (right) and B (left) with a resolution of 0.5~mas. 
The contour levels for image B are 1.5~mJy/beam $\times$ (--1, 1,
2, 4, 8, 16, 32, 64, 128) and the peak flux density  
is 130.8~mJy/beam.
The contour levels for image A are 2.0~mJy/beam $\times$ (--1, 1,
2, 4, 16, 32, 64, 128, 256) and the peak flux density is
381.4~mJy/beam. The
restoring beam of size 0.5~mas is drawn in the bottom left hand corner of
each panel. The tick interval is 1 mas.  Component 1 is the western
component in each image.}
\end{figure*}

At resolutions around a few milliarcsec we are not sensitive to
extended emission from the Einstein ring.  At resolutions higher than
about 1~mas, which is achieved for frequencies higher than 8.4~GHz,
each of the lensed images, A and B, is resolved into a `core-jet'
structure (Fig. 1).  We identify A1 (the south-western component of A)
and its lensed counterpart B1 (the western component of B) as the core
of the background radio source. Correspondingly, A2 and B2 are
identified as part of the jet. This identification is justified from
the fact that A1 and B1 have flatter spectra than A2 and B2
(Table 1, Fig. 2) and
also from the fact that they are
unresolved at 43~GHz with a resolution of 0.2~mas.

The images A and B have diffuse structure at 1.7 and 5~GHz such that
it is difficult match corresponding features in them. Image A shows
characteristic `tangential' elongation in PA $-$40$^{\circ}$. This
elongation is also seen at higher frequencies.  The other
noticeable feature is the rather dramatic increase in size of image A
and B with wavelength.  Even though the sizes of the images increase
with decreasing frequency, their surface brightnesses at low
levels are the same as expected from gravitational lensing.

\begin{table}
\caption{Flux densities (in mJy) of components in B0218+357.} \label{tbl-1}
\begin{center}\scriptsize
\begin{tabular}{llrrrrrrr}
Frequency & Date & Resolution &     &      &      &     &    &    \\ 
          &      & in mas     & A   &  B   &   A1 & B1  & A2 & B2 \\
         &              &   &     &     &   &   &   &    \\
1.7GHz   &  1992 Jun 19 & 5 & 445 & 170 & - & - & - & -  \\
5.0GHz   &  1992 Mar 27 & 1 & 515 & 196 & - & - & - & - \\
8.4GHz   &  1995 May 09 & 1 &  -  & -  & 472 & 139 & 218 & 63 \\
15 GHz   &  1995 Jul 17 & 0.5 & - & -  & 450 & 295 & 121 & 66 \\
22GHz    &  1995 Jul 17 & 0.3 & - &  - & 328 & 140 & 100 & 41 \\
43GHz    &  1995 Jul 17 & 0.2 & - &  - & 270 & 60  & 75  & 22 \\

\end{tabular}
\end{center}
\end{table}

Table 1 summarises the results of the multi-frequency VLBI observations.
The flux densities of various components (identified in Fig.1) are
plotted in Fig.2.

\begin{figure}
\hspace{3cm}
\raisebox{+3.0cm}
{\begin{minipage}{0.3cm}
\mbox{}
\parbox{0.3cm}{}
\end{minipage}}
\begin{minipage}{11.0cm}
\mbox{}
\centering
{\psfig{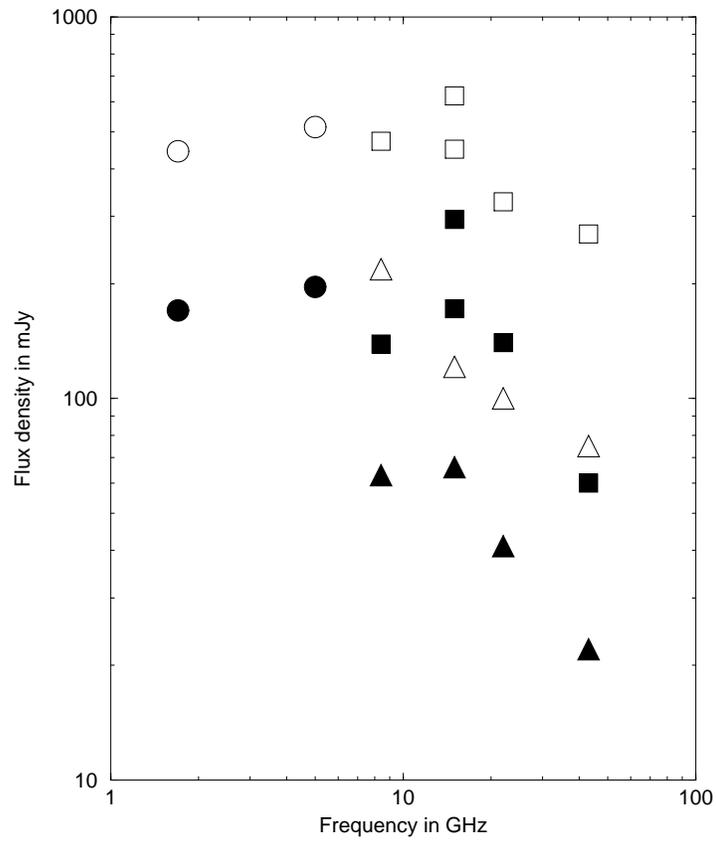}}
\centering
\end{minipage}
\vspace {1.5cm}
\caption{Spectrum of B0218+357. The various symbols are: open circle:
A; open square: A1; open triangle: A2; filled circle: B; filled
square: B1; filled triangle: B2. The errors on the flux densities are
expected to be about 5\%. However, the flux ratios are more accurate.}

\end{figure}

\section{Discussion}

We derive two basic results from these observations, namely the
spectral decomposition of various components and the ratio of their flux
densities as a function of frequency.

Since we resolve the `core-jet' structure in A and B at frequencies
higher than 8.4~GHz, we have 4 measurements of flux density for these
components. The spectra are plotted in Fig. 2. 
The spectrum of the core (A1 and B1) is self-absorbed with a peak
around 15~GHz. The jet (A2 and B2) has steep
spectra above 8.4~GHz. However, the total flux density of A and B at
1.7 and 5~GHz suggest that both A2 and B2 must also be self-absorbed
between these two frequencies.
The Einstein ring emission has a steeper
spectrum between 5 and 22~GHz and
it must also be self-absorbed at lower frequencies since
the total flux density provides a limit (Patnaik et al. 1993).

In summary, the radio structure of B0218+357 consists of at least 3
distinct components, core, jet and the Einstein ring, which are all
self-absorbed at different frequencies. The components are
self-absorbed at progressively lower frequencies as one moves from the
core along the jet. This behaviour of the radio source is consistent
with its flat spectrum (Cotton et al. 1980).

The core, being self-absorbed around 15~GHz, contributes very little
to the emission at 1.7 and 5~GHz, where the
emission is dominated by the jet as evident from the spectra.
Perhaps it is not surprising therefore that the radio structures of
both A and B at 1.7~GHz appear diffuse. However, this poses a
difficulty in that a self-absorbed component, which must be compact, has
a large observed size. In this
case one must consider the intrinsic size of the source rather than
the observed size which has been magnified by the lens.

Another result from our observations is that the image flux ratios
change with frequency. This is an apparent
contradiction since gravitational lensing is achromatic.
However, since the source
is extended and spans areas of different image magnifications the
observed flux ratio can indeed vary with frequency.
The flux ratio of the core (i.e. A1/B1) is around 3.7 for frequencies higher
than 8.4~GHz. This is similar to the ratio of A/B obtained from VLA
observations at these frequencies (Patnaik et al. 1993).   
At lower frequencies the flux ratio A/B is around 2.6.  Of
course, we point out that the source is variable at
high frequencies and thus the measured flux ratio can be different
due to differential time delay.

The above results lead to a rather simple model for B0218+357. The radio
source consists of a number of different components which are self-absorbed
at different frequencies thus conspiring to produce the observed flat spectrum.
The core (imaged into A1 and B1) is located at the western-most edge,
the jet (imaged into A2 and B2) lies between the core and the
component giving rise to the Einstein ring. The core dominates the
radio structure at frequencies higher than 8.4~GHz, and the jet and
the Einstein ring
dominate at lower frequencies.
At lower frequencies one does not detect any
compact feature in the source since the core contributes very
little to the flux density.

\acknowledgments

We would like to thank Athol Kemball for allowing us to quote results
before publication and Karl Menten for his comments.  The Very Long
Baseline Array of NRAO is operated by AUI under cooperative agreement
with the NSF, USA.

\end{document}